\documentclass{emulateapj}

\usepackage{graphicx}
\usepackage{psfig}
\usepackage{amsmath}

\def \nh {N${\rm _H}$}

\def \arcmin {\hbox{$^\prime$}}
\def \arcsec {\hbox{$^{\prime\prime}$}}

\def\spose#1{\hbox to 0pt{#1\hss}}
\def\ltsim{$\mathrel{\spose{\lower 3pt\hbox{$\sim$}}
        \raise 2.0pt\hbox{$<$}}$\thinspace}
\def\gtsim{$\mathrel{\spose{\lower 3pt\hbox{$\sim$}}
        \raise 2.0pt\hbox{$>$}}$\thinspace}
\def \msun {${\rm M_\odot}$}

\newcommand\solar{\hbox{{$Z_{\odot}$}}}

\def \nh {$N_{\rm H}$}

\newcommand{\source}{\mbox{Zw\,1305.4+2941}}

\newcommand{\apec}{APEC}
\newcommand{\mekal}{MEKAL}

\newcommand{\chandra }{{\em Chandra}}
\newcommand{\xspec }{{\em Xspec}}

\newcommand{\fxunits}{\mbox{ergs cm$^{-2}$ s$^{-1}$}}
\newcommand{\lxunits}{\mbox{ergs s$^{-1}$}}
\newcommand{\xmm }{{\em XMM}}
\newcommand{\asca }{{\em ASCA}}
\newcommand{\rosat }{{\em ROSAT}}

\newcommand{\tx }{${\rm T_X}$}

\newcommand{\ned}{{\em{NED}}}

\newcommand\omegam{\hbox{{$\Omega_{\rm m}$}}}
\newcommand\omegalambda{\hbox{{$\Omega_{\Lambda}$}}}
\newcommand\kmsmpc{{\rm km s$^{-1}$ Mpc$^{-1}$}}
\newcommand\ho{\hbox{{$H_{0}$}}}
\slugcomment{Submitted Aug 1, 2007; Accepted Sep 19, 2007 for publication in the Astrophysical Journal }
\shorttitle{\xmm\ observation of the cluster ZW 1305.4+2941}
\shortauthors{Gastaldello et~al.}

\begin{document}




\newcommand{\lessim}{\ \raise -2.truept\hbox{\rlap{\hbox{$\sim$}}\raise5.truept
    \hbox{$<$}\ }}

\title{{\em XMM-Newton} observation of the cluster ZW 1305.4+2941 in the field SA 57}

\author {Fabio Gastaldello\altaffilmark{1,2},
         Dario Trevese\altaffilmark{3},
         Fausto Vagnetti\altaffilmark{4}, 
         Roberto Fusco-Femiano\altaffilmark{5}
}
\altaffiltext{1}{Department of Physics and Astronomy, University of
California at Irvine, 4129
Frederick Reines Hall, Irvine, CA 92697-4575}
\altaffiltext{2}{Dipartimento di Astronomia, Universit\`a di Bologna, via
Ranzani 1, Bologna 40127, Italy}
\altaffiltext{3}{Dipartimento di Fisica, Universit\`a di Roma ``La Sapienza'', 
P.le A. Moro 2, Rome 00185, Italy}
\altaffiltext{4}{Dipartimento di Fisica, Universit\`a di Roma ``Tor Vergata'', 
via della Ricerca Scientifica 1, Rome 00133, Italy}
\altaffiltext{5}{Istituto di Astrofisica Spaziale e Fisica Cosmica (IASF/Roma), INAF, Rome, Italy}
\begin{abstract}
We report the details of an \xmm\ observation of the cluster of 
galaxies \source\ at the intermediate redshift of $z$=0.241, increasing 
the small number of interesting X-ray constraints on properties of $\sim3$ keV 
systems above $z$=0.1.
Based on the $\sim45$\,ks \xmm~observation, we find that within a radius of 
228 kpc the cluster has an unabsorbed 
\mbox{X-ray} flux of $f_{\rm X}=(2.07 \pm 0.06) \times 10^{-13}$ \fxunits, a 
temperature of $kT=3.17\pm0.19$\,keV, in good agreement with the previous 
\rosat\ determination, and an abundance of $0.93^{+0.24}_{-0.29}$ \solar.
Within $r_{500}=723\pm16$ kpc the rest-frame bolometric \mbox{X-ray} 
luminosity is $L_{\rm X}(r_{500})=(1.25 \pm 0.16) \times 10^{44}~h_{70}^{-2}$ \lxunits. The cluster obeys the scaling relations for $L_{\rm X}$, $T$ and 
the velocity dispersion $\sigma_v$ derived at intermediate redshift for 
$kT$\ltsim4 keV, for which we provide new fits for all literature objects. 
The mass derived from an isothermal NFW model fit is,
$M_{\rm vir} = 2.77\pm0.21 \times10^{14}$ \msun, with a concentration
parameter, $c = 7.9\pm0.5$.
\end{abstract}

\keywords{galaxies: clusters: general --- \mbox{X-ray}s: general}

\maketitle
\section{Introduction}

The current generation of X-ray observatories, \chandra\ and \xmm, is
considerably extending the maximum redshift to which X-ray clusters
can be identified and ana\-ly\-zed. In fact the number of massive clusters
detected at $z > 1$ is rapidly growing thanks to the unprecedented \xmm\
sensitivity \citep[e.g,][]{mullis05,stanford06,bremer06}. On the other 
hand they are also extending the minimum luminosity, i.e. the least massive
structures, to which X-ray clusters can be detected and analyzed at
intermediate redshifts. Galaxy groups and clusters with $kT$\ltsim 4 keV are
starting to be routinely detected and analyzed in detail at $0.2 < z < 0.6$
\citep{willis05,gaga05,jeltema06,puccetti06}, where few examples were known. 
They represent the population which \chandra\ and in 
particular \xmm\ surveys \citep[like the \xmm-LSS,][]{pierre04} are 
sampling using typical exposures (10-20 ks), as expected \citep{jones02}.

These objects are more likely to display the effects of
non-gravitational energy into the intra-cluster medium (ICM) than hotter
more massive clusters \citep[e.g.,][]{ponm03}. The study of X-ray extended
objects over an extended temperature range at $z> 0.2$ will provide an
important insight into the evolution of their X-ray emitting gas and the
deviation of X-ray scaling relations from simple, self-similar expectations.
Studies of objects in the redshift range $0.2 < z < 0.6$ and with 2 keV $< T <$
2.6 keV are already suggesting that at these redshift these objects are less
dynamically evolved that their counterparts at $z=0$ \citep{mulch06}. 
Furthermore, clusters with masses in the range $3-5\times10^{14}$ \msun\
(3 keV $<$\tx$<$ 4 keV) will constitute the bulk of the cosmological
constraining power of future SZ surveys, because this is the range well-above 
the nearly redshift independent detection limit of these surveys
\citep{haiman01}, therefore constituting the largest population in
number count studies.

Here we present details of the \xmm\ observation of the cluster \source, also
known as MS 1305+29 with $kT\sim3$ keV at the redshift of $z=0.241$, observed
during an exposure of the field SA57. All
distance-dependent quantities have been computed assuming \ho = 70 \kmsmpc,
\omegam = 0.3 and \omegalambda = 0.7. At the redshift of $z=0.241$ 1\arcmin\
corresponds to 228 kpc. All the errors quoted are at the 68\% confidence
limit.

\section{X-ray Analysis}

\begin{figure*}[th]
\begin{center} \includegraphics[width=0.5\textwidth]{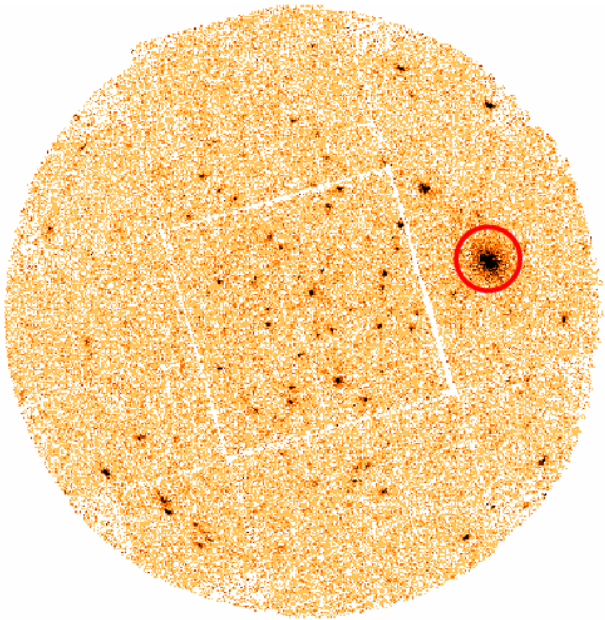}
\caption{Exposure corrected 0.5-2.0 keV combined MOS1 and MOS 2 X-ray
image of the field SA 57. The cluster \source\
is clearly visible as the extended source and it is highlighted by the red 
circle of 1.5\arcmin\ radius.}
\label{fig.1} \end{center}
\end{figure*}

The object \source\ has been observed during an \xmm\ pointing 
of the field SA 57 \citep{trevese07} and it is located at 9.3\arcmin\ off-axis 
(see Fig.\ref{fig.1})\footnote{The center of the pointing has been calculated 
as an effective-area-weighted average of the optical axis of the three 
telescopes taken form the exposure map headers keywords XCEN and YCEN, as in 
\citet{pacaud06}}.
The data were reduced with SAS v7.0.0 using the tasks
{\em emchain} and {\em epchain}. We considered only event patterns
0-12 for MOS and 0 for pn and the data were cleaned using the standard
procedures for bright pixels and hot columns removal (by applying the
expression FLAG == 0) and pn out-of-time correction.
Periods of high backgrounds due to soft protons were filtered as in
\citet{gasta07a}; the observation was affected by flares
at the end of the exposure and $\sim 20$ ks were lost, resulting in a net 
exposure time of 47, 48 and 40 ks respectively for MOS1, MOS2 and pn.

\begin{figure*}[th]
\begin{center} \includegraphics[width=0.4\textwidth,angle=-90]{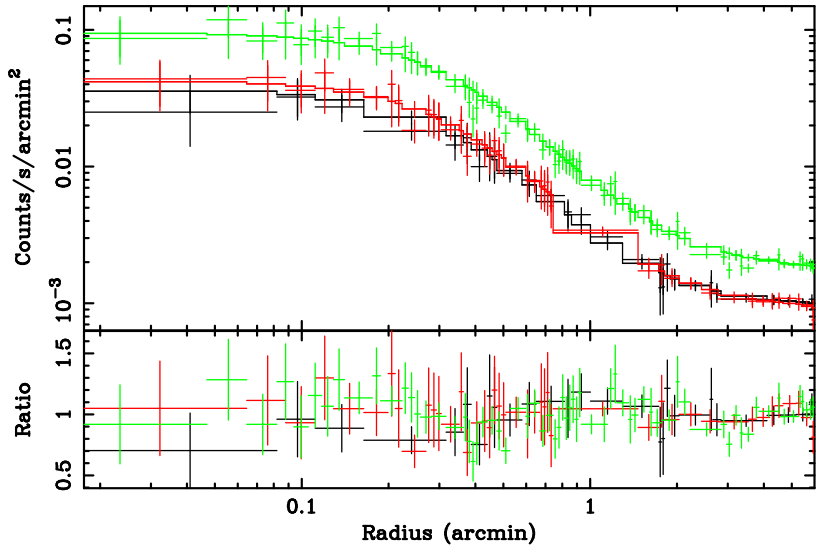}
\caption{Surface brightness profile of the X-ray emission of \source. 
Data from MOS1, MOS2 and pn are plotted in black, red and green respectively. 
The best fit beta model and ratio of data over the model are also shown.}
\label{fig.2} \end{center}
\end{figure*}

For each detector we created images in the 0.5-2 keV band with point sources,
detected using the task {\em ewavelet}, masked using circular
regions of 25\arcsec\ radius centered at the source position. The images have
been exposure corrected and a radial surface brightness profile was extracted
from a circular region of 6\arcmin\ of radius centered on the cluster
centroid. We account for the X-ray background by including a 
constant-background component. The data were grouped to
have at least 20 counts per bin in order to apply the $\chi^{2}$ statistic.
The fitted model is convolved with the \xmm\ PSF. The joint best-fit 
$\beta$-model \citep{beta} has a core radius of $r_c = 57\pm8$ kpc 
($14.9$\arcsec$\pm2.0$\arcsec) and $\beta=0.54\pm0.02$ for a 
$\chi^{2}$/d.o.f. = 202/130 (see Fig.\ref{fig.2}). Fits to the profiles of 
the individual detectors give consistent results within 1$\sigma$ of the 
combined-fit result and in the case of the MOS detectors are formally 
acceptable (20/24 MOS 1 and 41/39 MOS 2). The main contribution to the
$\chi^{2}$ comes mainly from the pn and its origin is instrumental, hence
there is no need for more complicated models.

\begin{figure*}[t]
  \begin{center} \includegraphics[width=0.4\textwidth,
  angle=-90]{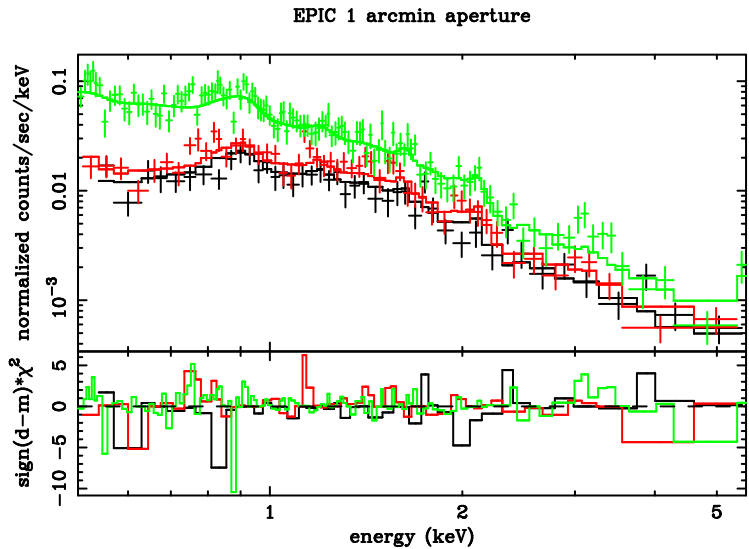} \caption{X-ray spectrum of the \source\ 
taken from a 1\arcmin\ aperture centred on the centroid of the emission. Data 
from MOS1, MOS2 and pn are plotted in black, red and green respectively. The 
best fit model and residuals are also shown.}
  \label{fig.3} \end{center}
\end{figure*}

For spectral fitting, we extracted spectra for each detector from a 1\arcmin\
region centered on the centroid of the emission, to maximize the S/N over the
background. Redistribution matrix files (RMFs) and
ancillary response files (ARFs) were generated using the SAS tasks
{\em rmfgen} and {\em arfgen} in extended source
mode. Appropriate flux-weighting was performed for RMFs and for ARFs, using 
exposure-corrected images of
the source as detector maps (with pixel size of 1\arcmin, the minimum
scale modeled by {\em arfgen}) to sample the variation in emission,
following the prescription of \citet{saxton02}. The background was estimated
locally using spectra extracted from a 2\arcmin-3\arcmin\ annular region
centered on the centroid of the emission. The spectra from the three 
detectors were re-binned to ensure a signal-to-noise
ratio of at least 3 and a minimum 20 counts per bin and they were
jointly fitted with with an \apec\ thermal plasma modified by Galactic
absorption \citep{dick90}. The spectral fitting was performed with
\xspec\ \citep[ver11.3.1,][]{xspec} in the 0.5-6 keV band and quoted 
metallicities are relative to
the abundances of \citet{grsa98}. The spectra are shown in Fig.\ref{fig.3}:
the best fit parameters are $kT = 3.17\pm0.19$ keV and  
$Z=0.93^{+0.24}_{-0.29}$ \solar\ for a $\chi^{2}$/d.o.f. = 241/222.

Using the best-fit model, the unabsorbed flux within the aperture of radius 
1\arcmin\ (228 kpc) is $2.07\pm0.06\times10^{-13}$ \fxunits\  in
the 0.5-2 keV band. This corresponds to an unabsorbed luminosity of
$3.33\pm0.15\times10^{43}$ \lxunits\ in the 0.5-2 keV and to a bolometric 
(0.01-100 keV) of $8.86\pm0.98\times10^{43}$ \lxunits. 
The quoted errors on
flux and luminosity are the mean and standard deviation of the distributions 
evaluated repeating the measurements after 10000
random selection of temperature, metallicity and normalization, drawn from 
Gaussian distributions with mean and standard deviation in accordance to the 
best fit results.

To investigate possible spatial variation in
the spectral parameters of the cluster, we extracted two annular
regions of radii 0\arcmin-0.5\arcmin\ and 0.5\arcmin-1.5\arcmin. The
derived spectral parameters are: $kT = 3.04\pm0.23$ keV and $Z=0.93\pm0.45$ 
\solar\ with $\chi^{2}$/d.o.f. = 140/121 for the inner annulus; 
$kT=3.47^{+0.22}_{-0.22}$ keV and $Z=0.91^{+0.32}_{-0.25}$ \solar\ with 
$\chi^{2}$/d.o.f. = 153/159 for the outer annulus. 
The width of the bins have been chosen in order to avoid bias in the 
temperature measurement caused by scattered flux by the PSF (80\% encircled 
energy fraction radius is 31\arcsec\ for the pn at 1.5 keV and at the 
off-axis angle of the source). The cluster is therefore 
consistent with being isothermal over the explored radial range.

The cluster has regular X-ray isophotes and is centered on a dominant early
type galaxy (see \S\ref{optical}). These characteristics suggest the cluster
is relaxed and that hydrostatic equilibrium is a good approximation.
The isothermal profile is not exceptional in a relaxed cluster, as 
\source\ seems to have properties very similar to the low redshift cluster 
A 2589 \citep{zappacosta06}.
We calculated the total mass profile using two different models.
First, we used the best-fit $\beta$-model for which
the gas density and total mass profiles can be expressed by simple
analytical formula \citep[e.g.,][]{ettori00}.
We evaluated $r_{500}$ as the radius at which the density is 500 times the
the critical density and the virial radius as the radius at which the
density corresponds to $\Delta_{\rm{vir}}$, as obtained by \citet{bryan98}
\footnote{$\Delta_{vir}=18\pi^2+82x-39x^2$ where $x=\Omega(z)-1$, $\Omega(z)=\Omega_0(1+z)^3/E(z)^2$ and $E(z)=\left[\Omega_m(1+z)^3+\Lambda\right]^{1/2}$} 
for the concordance cosmological model used in this paper.
To evaluate the errors on the estimated quantities,
we repeat the measurements after 10000 random selections of a temperature and
parameters of the surface brightness profile, which were drawn from Gaussian
distributions with mean and variance in accordance with the best-fit results.
For $\Delta=500$ we obtained,
$M_{500} = (1.37\pm0.15) \times 10^{14}$ \msun\ within
$r_{500} = 723\pm16$ kpc;
the virial mass is, $M_{\rm{vir}}= (2.81\pm0.30)\times10^{14}$ \msun,
within the virial radius $r_{\rm{vir}} = 1474\pm52$ kpc.
Secondly, we fit the surface brightness profile
with an isothermal NFW \citep{nfw} model \citep{suto98}.
We obtain a concentration parameter,
$c=7.9\pm0.5$, virial radius $r_{\rm{vir}} = 1468^{+34}_{-41}$ kpc,
and virial mass $M_{\rm{vir}}=(2.77\pm0.21) \times10^{14}$ \msun, with
$\chi^{2}$/d.o.f. = 197/130. The mass determinations using the two different
models agree well within the $1\sigma$ errors.
We calculate the gas mass using the procedure described in \citet{ettori04},
using the $\beta$-model parametrization and deriving the central electron 
density from a combination of the surface brightness fit and the 
normalization of the spectral model \citep[equation (2) of][]{ettori04}.
We obtained $M_{gas,500} = (1.37\pm0.14) \times 10^{13}$ \msun. 
We calculate the entropy of the cluster using the standard definition
$S=T_{gas}/n_e^{2/3}$ and measure this quantity at 0.1$r_{200}$ and $r_{500}$
as done by \citet{ponm03}. We find $S(0.1 r_{200}) = 168\pm12$ keV cm$^2$ and 
$S(r_{500}) = 1103\pm122$ keV cm$^2$, being $r_{200}=1144\pm41$ kpc using the 
best-fit $\beta$-model.

Finally, we have studied the sensitivity of our spectral results to various 
sources of systematic errors which we summarize below.

{\it Galactic Column Density and Bandwidth:} If \nh\ is allowed to vary the 
fit in the 1\arcmin\ aperture does not improve and the best fit column 
density is consistent at 1$\sigma$ with the Galactic value; the other 
parameters are unchanged. Restricting the energy band to the 0.5-5 keV band
returned practically unchanged values, $kT = 3.15\pm0.23$ and $Z=0.91\pm0.29$, 
whereas using a 0.4-5 keV band has the effect of slightly increase the 
values, $kT = 3.24\pm0.12$ and $Z=1.06^{+0.34}_{-0.12}$, but still with 
systematic errors less than the statistical ones.

{\it Background:} For comparison with the results obtained with the local 
background we used the standard blank background fields \citep{xmmbkg}, 
finding good agreement between the two methods in the 1\arcmin\ aperture, 
$kT = 3.20\pm0.12$ and $Z=0.87\pm0.23$. In the 0.5\arcmin-1.5\arcmin outer 
annulus the results obtained with the background template are in agreement 
within $1\sigma$ with the local background method, $kT = 3.41^{+0.25}_{-0.19}$ 
and $Z=0.70^{+0.22}_{-0.19}$.

{\it Plasma Code:} We investigated the sensitivity of our results to
the plasma code using the \mekal\ model. The quality of the fit and
the temperature and abundance values were found to be very consistent within 
the $1\sigma$ errors, $kT = 3.24\pm0.13$ and $Z=0.89^{+0.18}_{-0.10}$.

\section{Optical analysis}\label{optical}

\begin{figure*}[t]
\centering { \epsscale{0.7} \plotone{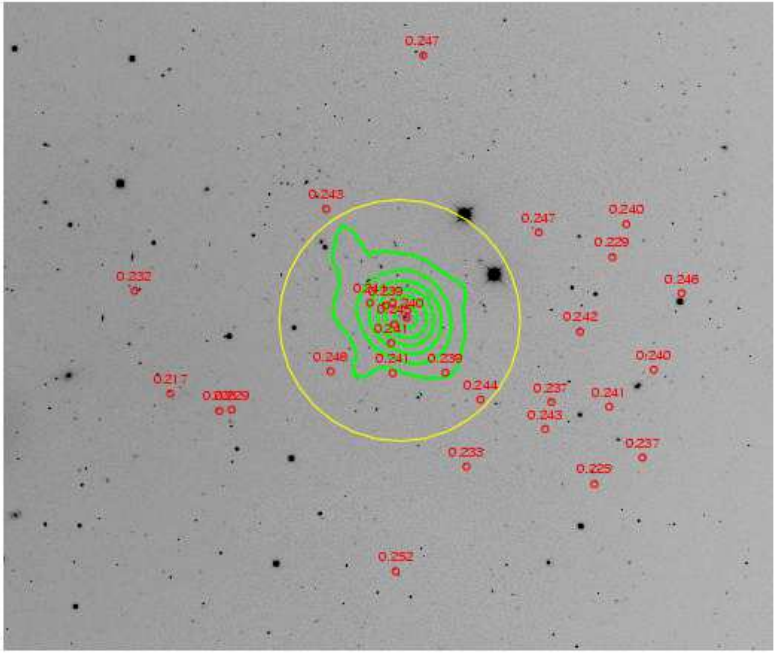}}
\caption{Optical map of the cluster ZwCl 1305.4+2941 with
over-plotted X-ray contours from XMM-Newton 0.5-10 keV data.
Galaxies in the redshift interval $0.21<z<0.27$ are indicated. The
yellow circle marks a region with radius 3\arcmin. \label{fig.4}}
\end{figure*}

Zw 1305.4+2941 is a Bautz-Morgan type I cluster at z=0.24. The
optical map with over-plotted X-ray contours from \xmm\ 0.5-10 keV
data is shown in Fig.\ref{fig.4}. Optical photometry of the field SA 57 was
obtained from U,$B_J$,F,N plates taken at KPNO 4 m Mayall
telescope \citep{koo86}. A foreground cluster was recognized
in the two-colour diagram calibrated with the spectroscopic
redshifts available in the field \citep{koo88}. The
cluster is elongated, with a major axis position angle $\theta_c
=57^{\circ} \pm 6^{\circ}$. The central cD galaxy has a position
angle $\theta_{cD}= 61^{\circ}$ and an axial-ratio $b/a = 0.78$,
both measured at 3.5 magnitudes fainter than the central surface
brightness and with a $\approx 10\%$ error. In \citet{koo88}
the galaxy number density profiles, as deduced respectively in
circular or elliptical annuli and statistically corrected for
background, were fitted with both a projected Emden isothermal
profile $\sigma(b)=\sigma_{o} F_{isot}(b/b_c)$ and projected King
profile $\sigma_{g}(b)=\sigma_{g_o}[1+(b/R_c)^2]^{-1}$ following
the procedure of \citet{sarazin80}. The parameters are the core
radius $R_c = 3b_c = 3\sigma_r/(4\pi Gn_0m)^{1/2}$, the central
surface density and the surface density of the background;
$\sigma_r$ is the dispersion of the radial velocities, $n_0$ the
central value of the volume density of galaxies and
\emph{\textit{m}} is the average galaxy mass. The results of the
fit are reported in Table 2B of \citet{koo86}.

\begin{figure}[th]
\centering { \epsscale{1.0} \plotone{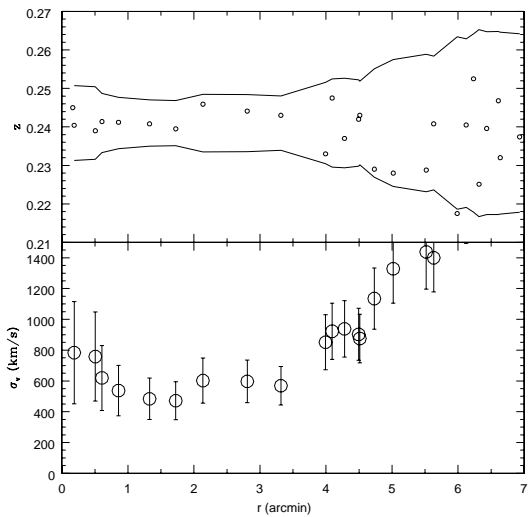}}
\caption{\textit{Upper panel}: redshifts of galaxies in the area
of ZwCl 1305.4+2941 as a function of the distance from the cluster
center. The lines represent $3\sigma$ limits around the cluster
redshift. \textit{Lower panel}: Velocity dispersion as a function
of the distance from the cluster center. Each error bar represent
the r.m.s. velocity dispersion computed by randomly extracting,
$10^4$ times, $N_{gal}(<r)$ velocities from gaussian distributions
with $\sigma_v(<r)$, where $N_{gal}(<r)$ and $\sigma_v(<r)$ are
the number and velocity dispersion of galaxies within distance $r$
from the cluster center. \label{fig.5}}
\end{figure}

\citet{mahdavi01} report a velocity dispersion value
$\sigma_r$ of $\sim 813~ {\rm km~s^{-1}}$ citing the paper of 
\citet{wu99}, but this value is not attributed in unequivocal way to
the cluster MS 1305.4+2941. \citet{mush97} used
velocity dispersion data from \citet{fadda96}, \citet{carlberg96} and 
\citet{fabricant91}, none of which contains
$\sigma_r$ for MS 1305.4+2941. On
the basis of the redshifts found using the Nasa Extragalactic
Database (\ned), mostly derived by the spectroscopic survey of the
field by \citet{munn97}, we have determined the velocity
dispersion of the cluster considering circular areas of increasing
radius, centered on the cD galaxy. The results are shown in Fig.\ref{fig.5}, 
where the redshifts as a function of the angular distance from
the cluster center are also reported. 
The velocity dispersion is roughly constant between 1 and 3.5\arcmin\ and 
progressively rises beyond 4\arcmin. This may be due to the inclusion 
of galaxies not belonging to the cluster or to incomplete virialization. 
The X-ray surface brightness falls below 3\% of the central value at a 
radius of 3.3\arcmin\ (see Fig.\ref{fig.2}), which corresponds to about 
$r_{500}$, and becomes practically undetectable. Therefore we adopt 
$\sigma(3.3^{\prime})=568\pm125$ km s$^{-1}$ (based on 10 members)
which corresponds to $\beta_{spec} \equiv \mu m_p
\sigma_r^2/kT= 0.64\pm 0.31$ for kT = 3.17 keV and $\mu$ = 0.6.
This value of $\beta_{spec}$ is not much different from the value
of $\beta_{fit}$ derived from the X-ray brightness distribution
($0.54\pm 0.02$). Moreover, $\beta_{spec} < 1$ is also consistent
with the X-ray image that in the central region seems to show a
quite relaxed cluster. A value of $\beta_{spec}$ lower than 1 may
be due to the transfer from orbital to internal energy occurring
in galaxy merging, thus \textit{cooling} the galaxy velocity
distribution despite the counteracting effect of the cluster
gravitational potential \citep{fusco95}. Finally,
the fraction of blue galaxies $f_b \sim 0.1$ derived within $1.4$\arcmin\ 
from \citet[][see their Fig.8]{koo88} for this cluster is
consistent with the relationship between cluster velocity
dispersion and blue fraction within $r_{200}$/4 obtained by
\citet{andreon06}, although not definitely conclusive.

\section{Discussion}\label{discussion}

\begin{figure}
 \begin{center} 
\includegraphics[width=0.35\textwidth,angle=-90]{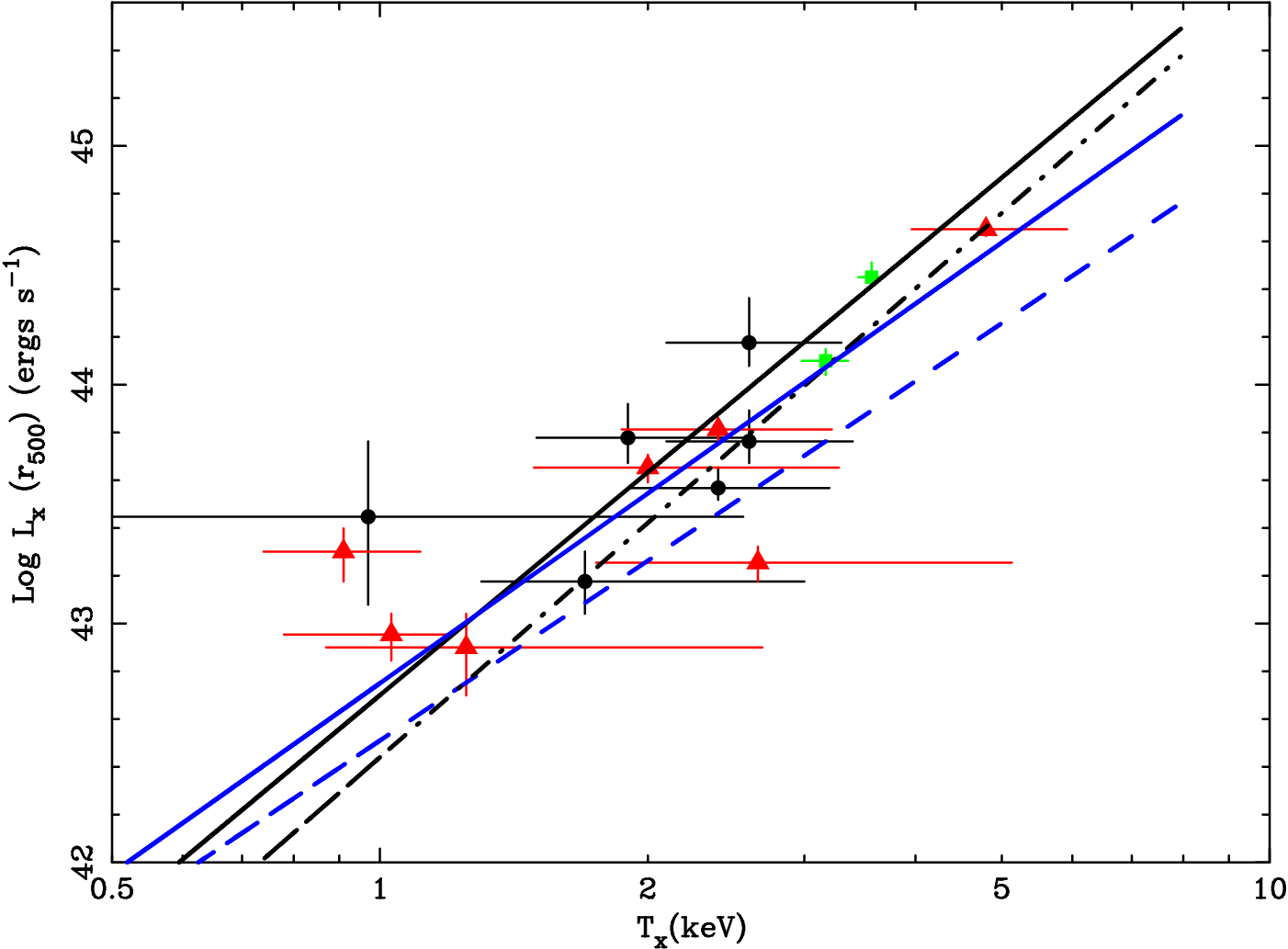}
\includegraphics[width=0.35\textwidth,angle=-90]{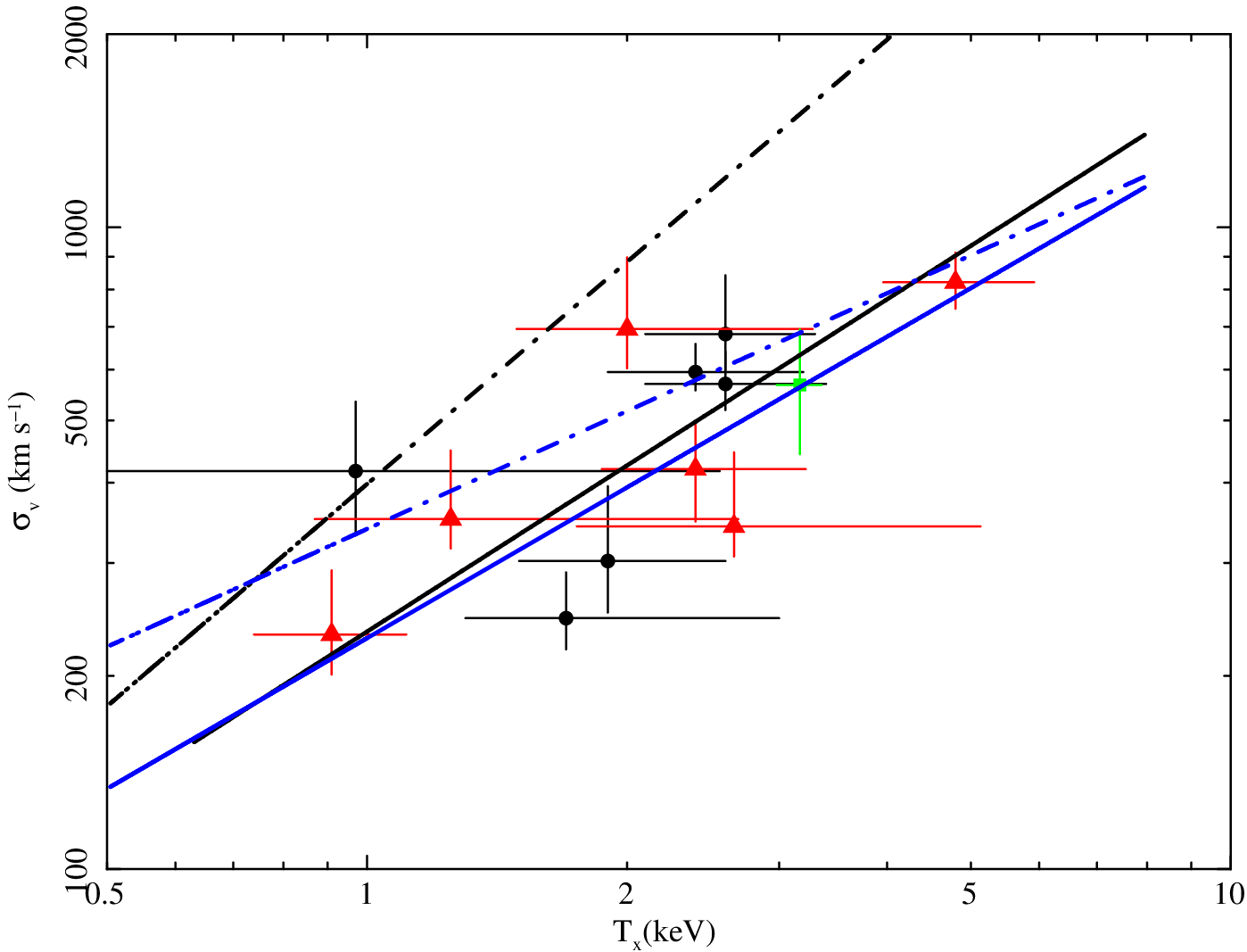}
\includegraphics[width=0.35\textwidth,angle=-90]{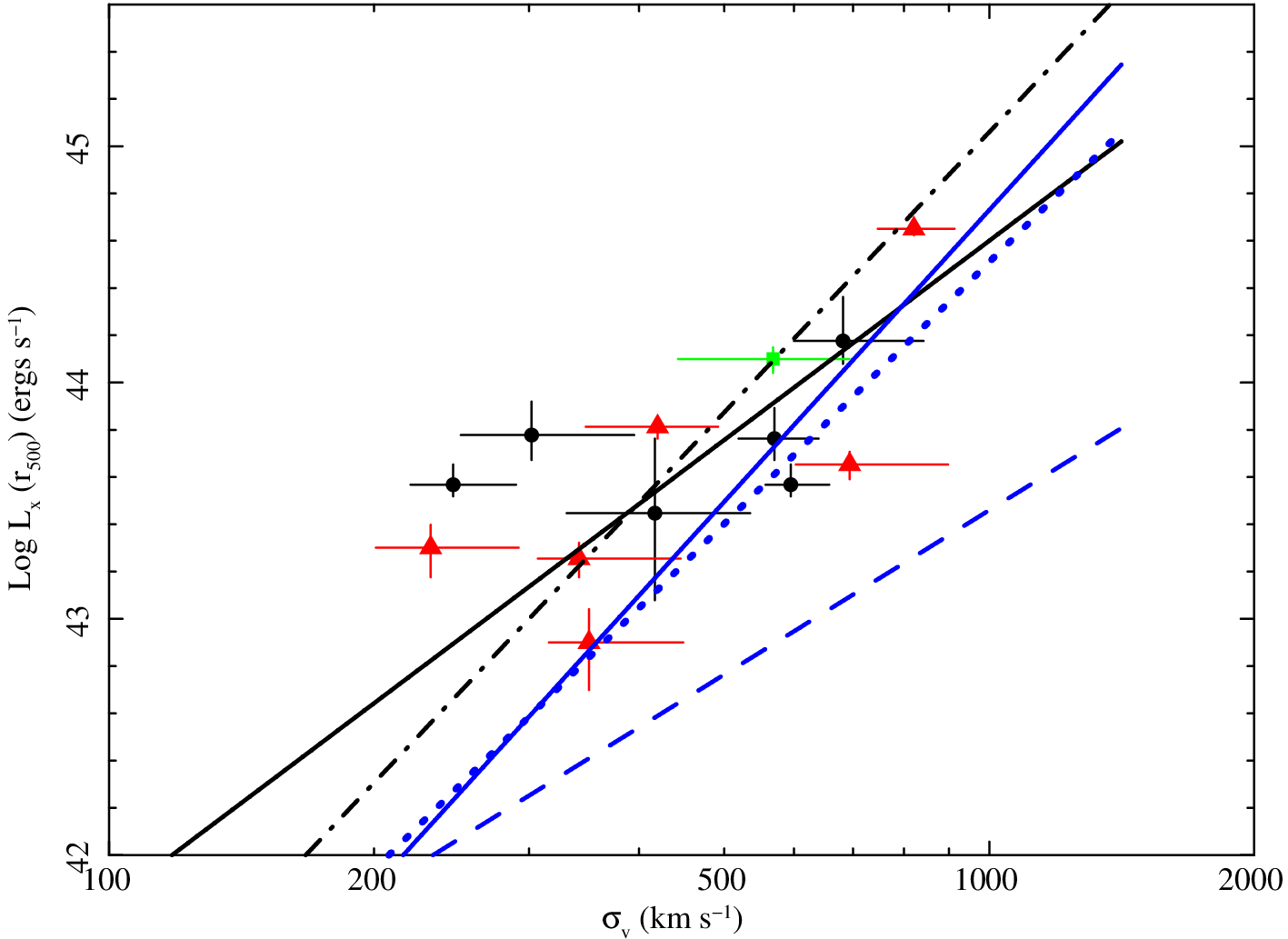}
\caption{\emph{Top panel:} $L_X-T_X$ relation for \source\ and the object discussed in 
\citet{gasta07b}(green squares), the moderate redshift sample of 
\citet{willis05}(red triangles) and the one of 
\citet{jeltema06}(black circles). The solid black line represents the best fit, the dot-dashed black line the fit to the cluster sample of \citet{horner01} as discussed in \citet{osmond04}, the solid blue line the fit to the clusters of \citet{mark98b} and the blue dashed line the fit to the groups of the GEM sample \citep{osmond04}.\newline
\emph{Central panel:} $\sigma_v-T_X$ relation with the best fit (black solid line). Also shown are the best fits to the cluster sample of \citet{horner01} (solid blue line) and \citet{girardi96} (dot-dashed blue line). The dot-dashed black line shows the best fit to the GEMS group sample. \newline
\emph{Lower panel:} $L_X-\sigma_v$ relation with the best fit (black solid line). Also shown are the best fits to the cluster sample of \citet{horner01} (dash-dotted black line), the REFLEX subsample of \citet{ortiz04} (solid blue line) and 
the RASS-SDSS sample of \citet{popesso05} (dotted blue line). The dashed blue line shows the best fit to the GEMS group sample.
\label{fig.6}}
\end{center}
\end{figure}

In this section we discuss the properties of \source\ in relation to 
the ones of objects with $kT$\ltsim4 keV at the intermediate redshift 
$0.2 < z < 0.6$.

In particular we investigate the scaling relations between
$L_x$, $T_X$ and velocity dispersion $\sigma_v$ for the six objects at 
intermediate redshift ($0.29<z<0.44$) in the \xmm\ LSS survey 
\citep{willis05}, the six objects 
($0.23<z<0.59$) in the sample of \citet{jeltema06} and \source\ together 
with the recently discovered cluster XMMUJ 131359.7-162735, with
$kT = 3.57\pm0.12$, presented in \citet{gasta07b}.
We made a first attempt at investigating quantitatively these relations 
(in a simple power-law representation) in its normalization and slope
by performing a linear regression $\log_{10} Y = \alpha \log_{10} X + b$
between two sets of measured quantities $Y$ and $X$. We employ the bisector 
modification of the BCES 
method (i.e., bivariate correlated errors with
intrinsic scatter) described by \citet{akritas96} that takes into account
both any intrinsic scatter and errors on the two variables considered as 
symmetric. We performed the BCES fitting using software kindly provided by
M. Bershady\footnote{http://www.astro.wisc.edu/$\sim$mab/archive/stats/stats.html}. The uncertainties on the best fit results are obtained from $10^{5}$ 
bootstrap re-sampling. The results on the best-fit normalization and slope 
for the scaling laws here investigated are quoted in Table \ref{tab.1}, 
together with total and intrinsic scatter 
\citep[measured using equation (2) of][]{buote07} and 
they are shown as black solid lines in Fig.\ref{fig.6}. 

We first investigate the $L_X-T_X$ relation. For \source, the aperture of 
228 kpc used for spectroscopy encloses 70\% of the flux within $r_{500}$, 
assuming the cluster emission profile
follows the $\beta$-model of Fig.\ref{fig.2}. The derived bolometric 
luminosity within $r_{500}$ is $L_{500} = (1.26\pm0.16) \times 10^{44}$ 
\lxunits; errors in the luminosity were determined including both the 
spectral errors and the uncertainties in the $\beta$-model parameters.
The temperature derived with \xmm\ is in good agreement with the previous 
\rosat\ determination 
\citep[$2.98^{+0.52}_{-0.41}$ keV, 2$\sigma$ errors,][]{mush97}.
In the first panel of Fig.\ref{fig.6} we plot the results for the 
intermediate redshift groups/poor clusters compared to the best fit 
regression lines for the low redshift groups of the GEMS sample 
\citep{osmond04}, for the clusters of \citet{horner01} as quoted in 
\citet{osmond04}, removing cool ($T_X<2$ keV) low luminosity 
($L_X < 2\times10^{43}$ \lxunits) objects and for the cluster sample of 
objects with  
$T_X>3$ keV of \citet{mark98b}. With the caveat of the large error bars in
the measured slope due to the still rather large errors in both luminosity 
and temperature and the small size of the sample, the relation is consistent
with what found for local clusters. Given the angular resolution of the data 
it has not 
been possible to correct for the effect of central cool cores, which tend 
to reduce the scatter and produce flatter slopes \citep{allen98c,mark98b}.

We then investigate the relationship between the velocity dispersion 
of the group member galaxies and the X-ray temperature, excluding from the 
sample of \citet{willis05} XLSSC 013 for which no 
velocity dispersion was quoted and XMMUJ 131359.7-162735 for which we do not 
have optical spectroscopy. For the sample of \citet{jeltema06} we used the 
updated velocity dispersions presented in \citet{jeltema07}. 
In the second panel of Fig.\ref{fig.6} we show the 
$\sigma_v-T_X$ relation compared to the best fit of the GEMS groups 
\citep{osmond04}, the cluster data of \citet{horner01} as quoted in 
\citet{osmond04} and the cluster sample of \citet{girardi96}. 
As discussed in \citet{jeltema06} there is a large scatter 
with few groups which appears to have significantly low velocity 
dispersions for their temperature, similar to what found in the GEMS 
sample at lower X-ray luminosities and temperatures. A well known 
observational effect could be a possible explanation, due to the fact that 
velocity dispersions may be artificially low when based on relatively small 
numbers \citep[e.g.,][]{zabludoff98,girardi01}. Or velocity dispersions could 
be really reduced: \citet{helsdon05} 
propose several possible mechanisms for this effect, including dynamical 
friction, tidal heating and orientation effect (see also discussion in 
\S\ref{optical}). The former explanation holds for many of the objects in the 
sample of \citet{jeltema06}: for example RX J1334.9+3750 increased its 
velocity dispersion from $121^{+58}_{-45}$ km s$^{-1}$ based on 6 members 
\citep{mulch06} to $246^{+44}_{-26}$ km s$^{-1}$ based on 17 members 
\citep{jeltema07}; RX J1648.7+6019  increased its velocity dispersion from 
$130^{+46}_{-48}$ km s$^{-1}$ based on 8 members to 
$417^{+118}_{-86}$ km s$^{-1}$ based on 22 members.
Deeper spectroscopy to increase the robustness of the determination of the 
velocity dispersion and deeper X-ray observations are therefore crucial to 
clarify the nature of these systems.
In the cluster regime there seems to be a consensus for a 
slight departure from a pure gravitational collapse, $\sigma_v \propto T^{\approx0.6}$ \citep[e.g.,][]{lubin93,bird95,girardi96,xue00}, whereas the evidence 
for groups is more controversial, with authors \citep[e.g.,][]{mulch00,xue00} finding that groups fall on the cluster trend and others \citep{helsdon00a,helsdon00b} finding that the relation steepens. As discussed in \citet{osmond04}
the large non-statistical scatter mentioned above contributes to the 
controversy. Again with
the caveat of large error bars, the objects in the intermediate-redshift 
sample seem to indicate 
an intermediate slope between clusters and low-temperature groups 
\citep[but see][for an even steeper slope, $1.00\pm0.16$, for a large REFLEX 
cluster sample]{ortiz04}. It is interesting to quote the results we would 
obtain had we used the velocity dispersions based on fewer members for the 
sample of \citet{jeltema06} reported in \citet{mulch06}: a slope of 
$0.89\pm1.18$ and intercept $2.30\pm0.18$.

Finally we investigate the $L_X-\sigma_v$ relation for the same objects 
considered above, shown in the third panel of Fig.\ref{fig.6}. The cluster 
relation slope is consistently found by many 
investigations close to the purely gravitational expectation of 4 
\citep[e.g.][see the latter reference for a thorough comparison with 
previous determinations]{girardi01,popesso05,ortiz04}. There is disagreement 
at the group scale between studies which find that groups are consistent 
with the cluster relation \citep[e.g.,][]{helsdon00a,mahdavi01} and those
which find significantly flatter relations \citep[e.g.,][]{xue00,osmond04}.
The results from the intermediate redshift sample point to a flatter trend 
compared to the cluster results. But we can see that this is mainly due to 
the same low velocity dispersion objects which have not only $T_X$ but also
$L_X$ higher compared to the expectations. Clearly these are not the 
X-ray under-luminous optically selected objects found both at the 
cluster \citep[e.g.,][]{popesso07a} and group scale 
\citep[e.g.,][]{rasmussen06} believed to be systems which are collapsing for 
the first time and it is therefore likely that a better determination of 
$\sigma_v$ in these X-ray selected objects will bring these objects in closer 
agreement to the cluster scaling relations 
\citep[modulo the effects described for example in][]{helsdon05}.
In fact, for example, had we used the velocity dispersions based on 
fewer members reported in \citet{mulch06}, we would have obtained a flatter 
slope of $2.07\pm1.21$ and intercept $38.3\pm3.2$.

We checked that the values for slope and intercept of the regression lines
do not depend on the choice of the pivot point: we fitted the relation
$\log_{10} Y/Y_0 = \alpha^{\prime} \log_{10} X/X_0 + b^{\prime}$ with
$L_0=10^{43.7}$ \lxunits, $T_0=2.0$ keV and $\sigma_0=300$ km/s, at the center of the data point cloud, in the three 
relations obtaining identical results to the ones in Table\ref{tab.1}.

\begin{table}[th]
\smallskip
\begin{center}
\caption{\label{tab.1} Best fit results on the scaling relations.}
\begin{tabular}{ccccc}
\tableline
\tableline
Relation ($Y-X$) & $\alpha$  & $b$ & $\sigma_Y$ & $\sigma_Y^{intr}$ \\
\tableline
$L_X-T_X$       &  $3.10\pm1.77$ & $42.7\pm0.6$  & 0.39 & 0.37 \\
$\sigma_v-T_X$  &  $0.86\pm0.85$ & $2.37\pm0.27$ & 0.14 & 0.12 \\
$L_X-\sigma_v$  &  $2.82\pm0.75$ & $36.2\pm2.0$  & 0.36 & 0.33  \\
\tableline\\
\end{tabular}
\tablecomments{Best fit results for the scaling relations discussed in the text. The total scatter $\sigma_Y$ on $Y$ is measured as 
$\sigma_Y=\left[ \sum_{j=1,N}\left(\log Y_j -\alpha -A \log X_j \right)^2 /N \right]^{1/2}$.
The intrinsic scatter is estimated as  $\sigma_Y^{intr}=\left(\sigma_Y^2-\sigma_{Y,stat}^{2}\right)^{1/2}$ where $\sigma_{Y,stat}^{2}=\sum_{j=1,N}\left(1/\sigma_{y_j}^2\right)/N$. 
}
\end{center}
\end{table}

\begin{figure}[th]
  \begin{center}\includegraphics[width=0.35\textwidth,
  angle=-90]{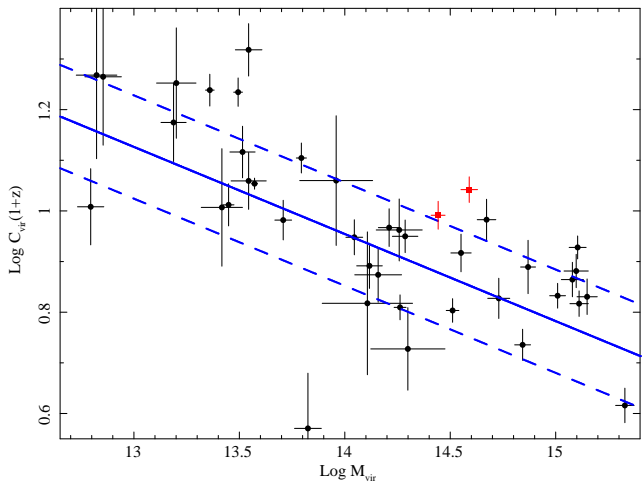} \caption{Local $c-M$ relation with data points (black circles) and best-fitting model (solid blue line) with $1\sigma$ intrinsic scatter (dashed blue lines) as discussed in \citet{buote07}. Over-plotted are the results for \source\ and XMMUJ 131359.7-162735 with red squares.}
  \label{fig.7} \end{center}
\end{figure} 

The entropy for \source, measured at the two different over-density of
0.1$r_{200}$ and $r_{500}$ is consistent with the excess entropy observed 
in low redshift groups: scaling our results as $E_z^{4/3}S$ with 
$E_z=H_z/H_0=\left[\Omega_m(1+z)^3+\Lambda\right]^{1/2}$ to account for the
variation of the mean density within a given over-density radius with redshift,
we find $E_z^{4/3}S(0.1 r_{200})=232\pm16$ keV cm$^2$ and $E_z^{4/3}S(r_{500})=1522\pm168$ keV cm$^2$ which can be compared with the plots in Fig.12 and 
Fig.13 of \citet{jeltema06} showing good agreement with the local relation

The high iron abundance for this cluster is consistent with the trend of
increasing metallicity with decreasing temperature found at intermediate and 
high redshift by \citet{balestra07} with \chandra\ and \xmm\ data and with the
sample of \citet{baumgartner05} with \asca\ data. Particularly relevant 
is the comparison with objects like Zw 0024.0+1652 and V1416+4446 in the 
sample of \citet{balestra07} which
at z=0.395 and z=0.400 have $kT = 4.38\pm0.27$, $Z=1.09^{+0.28}_{-0.26}$ and 
$kT = 3.50\pm0.18$, $Z=1.34^{+0.31}_{-0.24}$ respectively, having converted 
their abundances from \citet{asplund05} to \citet{grsa98} by scaling for 0.89.
Also these objects, observed with \chandra\ and for which an analysis in two
annuli is performed, do not show a clear enhancement of the iron abundance 
in the inner regions \citep{balestra07}, displaying the same behavior as 
\source. These data seem to suggest therefore that the higher abundance is 
not due to the presence of a particularly iron-rich cool core. 
\source\ is present in the sample of \citet{baumgartner05} based on the 
results of \citet{horner01}: the ASCA abundance determination is 
high ($0.92^{+0.86}_{-0.59}$, 90\% errors, having converted their abundances 
from \citet{angr} to \citet{grsa98} by scaling for 1.48) and in good agreement 
with the \xmm\ measurement.
This trend between 
metallicity and temperature, still poorly understood, needs to be 
investigated with better data (it should be remarked that the constraints on
the metallicity of \source\ are interesting but not tight: it is consistent 
with $\sim0.3$ \solar\ at the $2\sigma$ level).

The measured concentration parameter $c$, multiplied by the expected 
dependence $1+z$ \citep{bullock01} is consistent with a relaxed, early 
forming object in a $\Lambda$CDM model with $\sigma_8=0.9$ and with the 
observational results of relaxed, low-$z$ objects \citep{buote07}. 
In Fig.\ref{fig.7} we show the data points corresponding to \source\ and 
XMMUJ 131359.7-162735 over-plotted to the $c-M$ data points and best-fit 
relation discussed in \citet{buote07}. It should be kept in mind that these
two data points have been derived under a very simple and restrictive 
isothermal assumption, whereas all the low-$z$ data points  
have been derived with a detailed investigation of the density, temperature 
and abundance profiles.

\section{Conclusions}
\label{conclusion}

We present results for an \xmm\ observation of the cluster \source, for
which we derive $kT=3.17\pm0.19$\,keV, an abundance of $0.93^{+0.24}_{-0.29}$ \solar\ for an unabsorbed bolometric luminosity of $8.86\pm0.98\times10^{43}$ \lxunits\ within an aperture of 1\arcmin\ (228 kpc at $z=0.241$). Under the
assumption of isothermality and that the cluster follows the best-fit model 
to the surface brightness profile, we derive luminosity, entropy and mass
at various over-densities. We measure a velocity dispersion of 
$568\pm125$ km s$^{-1}$ within 3.3\arcmin. These interesting constraints
increase the small number of well studied $\sim3$ keV objects above $z$=0.1.

We provide new fits of scaling relations for all literature $kT$\ltsim4 keV 
objects in the intermediate-$z$ range. The cluster obeys the scaling 
relations thus derived. Concentration and mass for this 
object agree with the local $c-M$ relation.

The prospects for increasing the sample size and improving the description
presented here are promising.
\xmm\ and \chandra\ are dramatically expanding our
previous little knowledge of X-ray emitting low-temperature clusters and 
groups of galaxies beyond the present epoch. 
Surveys like \xmm-LSS \citep{pierre04} and 
COSMOS \citep{fino07} will provide large samples of X-ray selected groups and
poor clusters out to redshift of $z \sim 0.6$ or higher. Together with 
very large redshift surveys, optically selected groups in large 
quantities at moderate redshifts \citep[e.g.,][]{wilman05a,wilman05b} will
be obtained, characterizing in great detail this population and investigating 
some initial 
suggestion of group downsizing. More massive groups could be 
still in the process of virializing at intermediate redshift, while this 
process is restricted to much less luminous (and thus less massive) systems 
at present day \citep{mulch06}. X-ray follow-up with \chandra\ 
could allow to go beyond the simple isothermal beta model used so far in these
studies.

\begin{acknowledgements}
We would like to thank D.A. Buote and S. Ettori for useful discussions. 
P.J. Humphrey is 
thanked for the use of his surface brightness fitting code. 
Partial support for this work was provided by the ASI/INAF grants n. I/023/05/0
and n. I/088/06/0.
This work is based on observations obtained with \xmm\, an ESA science
mission with
instruments and contributions directly funded by ESA member states and
the USA (NASA).
This research has made use of data obtained from the
High Energy Astrophysics Science Archive Research Center (HEASARC),
provided by NASA's Goddard Space Flight Center.  This research has
also made use of the NASA/IPAC Extragalactic Database (\ned) which is
operated by the Jet Propulsion Laboratory, California Institute of
Technology, under contract with NASA. 
\end{acknowledgements}

\bibliographystyle{apj}
\bibliography{fgrefs}
\end{document}